\documentstyle[prl,aps,twocolumn,floats,amssymb,epsfig]{revtex}
\def\bbZ{{\mathbb Z}}

\def\SU{{\rm{SU}}}
\def\U{{\rm{U}}}
\newcommand{\mb}[1]{\ifmmode#1\else\mbox{$#1$}\fi}

\newcommand\ga{\mb{\gamma}}
\newcommand\de{\mb{\delta}}

\newcommand\la{\mb{\lambda}}
\newcommand\calC{\mb{{\cal C}}}

\newcommand\calE{\mb{{\cal E}}}

\newcommand\calL{\mb{{\cal L}}}

\newcommand\calP{\mb{{\cal P}}}

\newcommand{\beq}{\begin{equation}}
\newcommand{\eeq}{\end{equation}}
\newcommand{\nn}{\nonumber}
\newcommand{\bea}{\begin{eqnarray}}
\newcommand{\eea}{\end{eqnarray}}

\newcommand{\mod}[1]{\mid \!\! {#1} \!\! \mid}

\newcommand{\tr}{\mb{{\rm tr}\,}}

\newcommand{\x}{\mb{\times}}
\newcommand{\rhat}{\mb{\hat{\bm r}}}

\newcommand{\gsim}
{\raise.3ex\hbox{$\;>$\kern-.75em\lower1ex\hbox{$\sim$}$\:$}}
\newcommand{\lsim}
{\raise.3ex\hbox{$\;<$\kern-.75em\lower1ex\hbox{$\sim$}$\:$}}
\newcommand{\ts}{\textstyle}
\newcommand{\ds}{\displaystyle}
\newcommand{\half}{\mb{\ts \frac{1}{2}}}

\newcommand{\third}{\mb{\ts \frac{1}{3}}}
\newcommand{\twothird}{\mb{\ts \frac{2}{3}}}
\newcommand{\fifth}{\mb{\ts \frac{1}{5}}}
\newcommand{\fourfifth}{\mb{\ts \frac{4}{5}}}
\newcommand{\wt}[1]{\widetilde{#1}}
\newcommand{\bm}[1]{{\mbox{\boldmath $#1$}}}
\newcommand{\sm}{{\rm SM}}
\newcommand{\rmC}{{\rm C}}
\newcommand{\rmCp}{{\rm C'}}
\newcommand{\rmI}{{\rm I}}
\newcommand{\rmY}{{\rm Y}}
\newcommand{\minus}{{\mbox{$$-$$}}}

\begin{document}
\draft

%%%%%%%%%%%%       abstract and title page    %%%%%%%%%%%%%%%%%%%%%%

\twocolumn[\hsize\textwidth\columnwidth\hsize\csname @twocolumnfalse\endcsname
\title{Dualising the Dual Standard Model}
\author{Nathan\  F.\  Lepora\footnotemark}
%\address{101 Larkhall Rise, Clapham, England}
\date{February 7, 2001}
\maketitle

\begin{abstract}
We discuss how the dual standard model and the dualised standard model
are complementary theories. That is, how their implications have no
overlap, whilst together they explain most features of the standard model. 
To illustrate how these two theories might be combined we consider the dual 
standard model in a theta vacuum. Whilst there are issues to be considered, 
the dual standard model does then appear to become naturally 
dualised. This supports an origin of a dual formulation of the 
standard model through the properties of SU(5) solitons in a theta vacuum.
\end{abstract}
\pacs{pacs no.s.}]

%%%%%%%%%%%%%         the text   %%%%%%%%%%%%%%%%%%%%%%%%%%%%%%%%%%%

\section{Introduction}
\footnotetext[1]{\ email: n$\_$lepora@hotmail.com} 

Recently a remarkable correspondence has been discovered between the 
monopoles from Georgi-Glashow gauge unification and the observed 
elementary particles. 
Vachaspati found that the magnetic charges of the five stable
SU(5) monopoles have a one-to-one identification with the electric charges
of the five multiplets in one standard model generation~\cite{vach95}.
Motivated by this he conjectured that the elementary particles may originate as
solitons from SU(5) gauge unification.

A concrete way for examining this conjecture has been proposed by Liu and 
Vachaspati in the form of the {\em dual standard model}~\cite{vachdsm}. 
This relies on
the notion, familiar from electromagnetism, that electric particles can
also be described by monopoles in the dual gauge potential. In this sense
the dual standard model would be the dual description of the standard model,
with all of the elementary particles represented instead as monopoles.

By expressing the standard model in this dual formulation it is possible that
there may emerge features that are presently hidden within the usual
particle description. That is, a dual standard model
may uncover a hidden simplicity and regularity of form that could prove 
crucial to understanding the nature and origin of the standard model. 
Also possible is that new physics may have to be included to arrive at
a simple and consistent form.

This discovery of the SU(5) monopole-particle correspondence
strongly hints that a dual standard model should be formulated around the 
monopoles from gauge unification. In Vachaspati's words~\cite{vach95}:
{\em This correspondence suggests that perhaps unification should be based on
a magnetic SU(5) symmetry group with only a bosonic sector and the presently
observed fermions are really the monopoles of that theory.}
Much work still needs to be done on this proposal, but several encouraging
features do occur. For instance the incorporation of spin~\cite{vachspin} 
and a consistent picture of confinement~\cite{vachdsm,gold99}. 

In this paper we are concerned with the construction of the dual standard model
and whether it could be naturally dualised, 
in the sense of Chan and Tsou~\cite{chandual}.
In their {\em dualised standard model}~\cite{chandsm} 
(which should not be confused with the dual standard model) they interpret many
properties of the elementary particles as emerging from duality; for instance
three generations arise from just one generation of dyons. 
Remarkably this gives accurate estimations for both the masses and mixing 
angles of the elementary particles~\cite{chanckm}.  

A central point of this paper is that both the methodology and the conclusions
of the dual and dualised standard models appear to be
complementary to each other. That is, there is no overlap in their 
conclusions, whilst taking the two models together appears to explain most
observed features of the elementary particles. For this reason we examine
whether these two models could be considered together.

To illustrate how these models might be combined we investigate
the dual standard model in a theta vacuum. Whilst
there are issues to be considered, it appears that the initial 
assumptions of the dualised standard model can emerge. In this sense the dual 
standard model becomes naturally dualised. Also, giving further
corroboration, this calculation appears to explain 
the chirality assignments of the elementary fermions; a feature that 
cannot be derived in either of the original models.

If the above two models can be combined together in such a simple and natural
way then perhaps a very simple theory of particle and gauge
unification could ensue. Indeed it seems possible that 
{\em all features of a dual standard model could naturally emerge within the 
properties of SU(5) monopoles in a theta vacuum.}
As we have mentioned such a behaviour does seem to be occurring. However more
research is necessary to determine whether this can be fully realised. 

The composition of this paper is as follows. In sec.~(\ref{secstart}) 
we briefly discuss the dual and dualised standard models and how they relate
to each other. Then in sec.~(\ref{dual}) we discuss how the two models may
naturally combine in a theta vacuum. Finally in sec.~(\ref{conc})
we draw our conclusions.

Before starting we note that an alternative viewpoint for realising a
dual standard model has been presented by Vachaspati and Steer~\cite{steer}.

\section{Duality and the Standard Model}
\label{secstart}

In this section we quickly remind the reader of some results within
the dual standard model and the dualised standard model. This
discussion is also intended to clarify the complementary 
roles these theories presently take. 

\subsection{The Dual Standard Model}
\label{dsm}

The construction of a dual standard model is based around a Georgi-Glashow
unification~\cite{gg} of the standard model gauge symmetry within an SU(5) 
group~\footnote{Note that (\ref{su5}) relies on the
elementary particles forming representations of $\SU(3)\x\SU(2)\x\U(1)/\bbZ_6$;
as is implied by an observed $\bbZ_6$ relation between their colour,
isospin and hypercharge assignments~\cite{cs81}.}
\beq
\label{su5}
\SU(5)\rightarrow H_\sm = \SU(3)_\rmC\x\SU(2)_\rmI\x\U(1)_\rmY/\bbZ_6,
\eeq
which breaks via condensation~\cite{higgs} of an adjoint scalar field. 
This implies a spectrum of stable SU(5) monopoles, having various colours, 
isospins and hypercharges. Their magnetic charges are specified by the 
magnetic field 
\beq
\label{b}
\bm B \sim \frac{1}{2g}\frac{\rhat}{r^2}M,\ \ \ \ 
M=m_\rmC T_\rmC + m_\rmI T_\rmI + m_\rmY T_\rmY,
\eeq
with a suitable choice of generators, for instance,
\bea
T_\rmC ={\rm diag}
(\mbox{-}\ts\frac{1}{3},\mbox{-}\ts\frac{1}{3},\ts\frac{2}{3},0,0),\ \ \ 
T_\rmI = {\rm diag}(0,0,0,\mbox{-}1,1),\nn\\
T_\rmY = {\rm diag}(1,1,1,\mbox{-}\ts\frac{3}{2},\mbox{-}\ts\frac{3}{2}).
\hspace{3em}
\label{gen}
\eea
When the scalar masses
are much smaller than the gauge masses Gardner and Harvey showed there 
are five, topologically distinct, stable monopoles with magnetic charges 
forming the pattern~\cite{gard84}:

\begin{table}[h]
\caption{SU(5) monopole charges.}
\begin{eqnarray*}
\begin{array}{|c|c|ccc|c|}
\hline
{\rm topology\ }n &
{\rm diag}\ M &m_{\rm C}&m_{\rm I}&m_{\rm Y}&{\rm multiplet} \\
\hline
\vspace*{-0.35cm} &&&&& \\
1&(0,0,1,$-$1,0)& 1 & \half & \third & (u, d)_L  \\
\vspace*{-0.35cm} &&&&& \\
2&(0,1,1,$-$1,$-$1)& $-1$ & 0 & \twothird & \bar{d}_L  \\
\vspace*{-0.35cm} &&&&& \\
3&(1,1,1,$-$2,$-$1) & 0 & $-$\half & 1 & (\bar{\nu}, \bar{e})_R\\
\vspace*{-0.35cm} &&&&& \\
4&(1,1,2,$-$2,$-$2)& 1 & 0 & \ts\frac{4}{3} & u_R \\
\vspace*{-0.35cm} &&&&& \\
6&(2,2,2,$-$3,$-$3)& 0 & 0 & 2 & \bar{e}_L \\
\hline 
\end{array}
\end{eqnarray*}
\vspace*{-0.5cm}
\label{tab1}
\end{table}

Based upon this a dual standard model
could be constructed along the following lines:

\noindent (i)
First and foremost the magnetic charges in table~\ref{tab1} are identical to 
the electric charges in one standard model generation~\cite{vach95}. This 
suggests that one generation of standard model particles have a monopole 
description as solitons from a dual $\wt\SU(5)$ unification of the magnetic
gauge symmetry 
$\wt{H}_\sm = \wt\SU(3)_\rmC\x\wt\SU(2)_\rmI\x\wt\U(1)_\rmY/\bbZ_6$ in the 
dual standard model.

\noindent (ii)
To represent standard model fermions these solitons should have an
intrinsic one-half angular momentum. This can
be naturally achieved through the fermions from bosons 
effect~\cite{jackiw:spin,gold:spin}; from which the
dyons formed from combining SU(5) monopoles and quanta of a 
$\bm 5$ scalar field $H$ have the requisite angular momenta~\cite{vachspin}. 

\noindent (iii) 
Confinement is expressed through breaking dual colour
$\wt\SU(3)_\rmC\rightarrow\bbZ_3$~\cite{vachdsm,gold99}, which attaches
the appropriate monopoles to topological vortices.

\noindent (iv)
When normalising the generators (\ref{gen})
to tr$\,T^2=1$ the gauge-monopole couplings naturally scale
within the minimal coupling $g A_\mu^a T^a$. This suggests the dual 
standard model unifies when
$\ts\frac{1}{3}\,g_\rmC = g_\rmI = \surd\ts\frac{15}{2}\,g_\rmY$~\cite{meunif}.
Curiously such scaled coupling do unify, although the scale of unification
is rather low and could prove problematic:
 
\begin{figure}[h]
\begin{center}
\epsfxsize=20em \epsfysize=14em \epsfbox{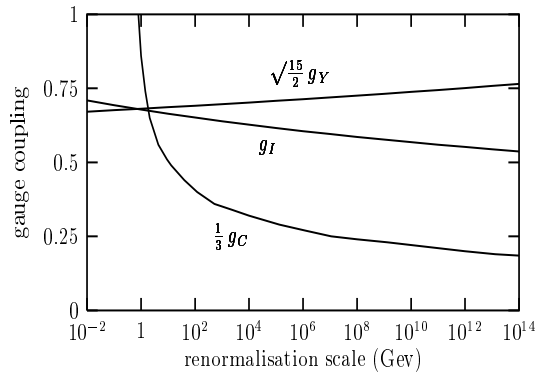}
\end{center}
\caption{Rescaled running gauge couplings.}
\label{fig1}
\end{figure}

\subsection{The Dualised Standard Model}

To construct a dualised standard model Chan and Tsou propose the
elementary fermions are dyonically charged, with dynamics depending
upon both an electric and magnetic gauge symmetry~\cite{chandual}
\beq
\label{dsd}
H_\sm \x \wt{H}_\sm.
\eeq
Note an independent treatment of Abelian dualised 
gauge symmetry has been given by Kleinert~\cite{kleinert}.

The main implication of the dualised structure (\ref{dsd}) is associated 
with the existence and properties of three standard model 
generations~\cite{chandsm}. Following a theorem of 't~Hooft, colour 
confinement implies dual colour $\wt\SU(3)_\rmC$ breaks to 
triviality~\cite{chanord}. Then a dual colour multiplet
\[
\psi = (\psi^{\wt r},\psi^{\wt g},\psi^{\wt b}),
\]
splits into three components, with each component's mass determined
by the details of the breaking. Interpreting this dual colour as a horizontal 
generational symmetry naturally leads to three generations of fermions from one
generation of dual colour charged particles. 

To describe symmetry breaking the relevant condensing scalar degrees 
of freedom must be identified. Chan and Tsou claim such scalar
fields occur as frame vectors within the non-Abelian electric-magnetic 
duality. In this sense these scalar fields are interpreted as independent 
degrees of freedom arising naturally from the dualised nature of (\ref{dsd}).
Within a dualised standard model this gives two isospin doublets 
and three dual colour triplets.

This structure allows an estimation of fermion masses by
constructing Yukawa couplings between the fermions and the dual colour
scalar fields. Analogous to electroweak theory this is possible
when only the isospin doublet fermions are dual colour charged. 
At tree level this coupling diagonalises into only one massive generation, 
which roughly approximates the standard model. To first order non-zero masses
are induced for the other two generations~\cite{chanckm}:

\begin{table}[h]
\caption{Particle mass predictions.}
\begin{eqnarray*}
\begin{array}{|c|c|c|}
\hline
& \ \ \ {\rm calculation}\ \ \  & \ \ \ {\rm experiment}\ \ \  \\
\hline
m_c & 1.327\, {\rm GeV} & 1.0-1.6\, {\rm GeV} \\
m_s & 173\, {\rm MeV} & 100-300\, {\rm MeV} \\
m_\mu & 106\, {\rm MeV} & 105.7\, {\rm MeV} \\
m_u & 235\, {\rm MeV} & 2-8\, {\rm MeV} \\
m_d & 17\, {\rm MeV} & 5-15\, {\rm MeV} \\
m_e & 7\, {\rm MeV} & 0.511\, {\rm MeV} \\
\hline 
\end{array}
\end{eqnarray*}
\vspace*{-0.5cm}
\end{table}

\noindent 
Note the poor match for the lightest generation, which they attribute
to their approximation techniques.
The CKM mixing angles also derive from the same inputs   
\[
\mod{V_{rs}} =
\left( 
\begin{array}{ccc} 
0.9752 & 0.2215 & 0.0048 \\
0.2210 & 0.9744 & 0.0401 \\
0.0136 & 0.0381 & 0.9992 
\end{array} \right);
\]
again these compare favourably with experiment
\[
\left( \begin{array}{ccc} 
0.9745 - 0.976 & 0.217 - 0.224 & 0.0018 - 0.0045 \\
0.217 - 0.224 & 0.9737 - 0.9753 & 0.036 - 0.042 \\
0.004 - 0.013 & 0.035 - 0.042 & 0.9991 - 0.9994
\end{array} \right).
\]

\subsection{Complementarity of the Dual and Dualised Standard Models}

In this section we discuss how the above two models complement each other.
That is, how their physical implications have no overlap, whilst their total
implications explain most of the standard model.
It is important to stress that these models are completely independent, 
and that they discuss different aspects of non-Abelian duality.

Firstly, the principle success of the dual standard model is to predict the 
electric charges for just one standard model generation, whilst it gives
no explanation for three generations. Complementary to this
the dualised standard model takes these electric charges as input, 
whilst deriving three generations.

Secondly, the dual standard model explains the origin of spin through
considering dyons instead of monopoles. Complementary to this the dualised
standard model takes these spins as input, whilst assuming the fermions are
dyonic to derive three generations. Later we will see that the specific 
representations required to achieve these effects can be consistent.

Thirdly, no particle masses have been derived in the dual standard model, 
whilst this is a central aspect of the dualised standard model.
Currently the only indication for the dual standard model mass scale is
through the gauge unification in fig.~\ref{fig1}, which suggests a few GeV.

Finally for electroweak symmetry breaking and confinement the dual standard
model assumes the necessary scalar field structure. 
Complementary to this the dualised standard model derives such
fields from the properties of non-Abelian duality.

We hope this gives some motivation for treating these two theories 
together. For further corroboration we now make some additional comments.

The point of both the dualised standard model and the dual standard model
is to express the standard model in a simpler form. The dualised standard
model does this by reducing the situation to essentially one generation
of fermions. In the dual standard model one generation of fermions is
understood to originate from gauge unification. In this sense the dual 
standard model 
reduces the fermions to simply a consequence of gauge interaction.

Finally the dual and the dualised standard model complement each other
on a theoretical level. The dual standard model is based on the notion that 
electric particles can also be described as monopoles in the dual 
gauge potential. The dualised standard model is based on a quite different 
aspect of duality, where both electric and magnetic interactions are 
considered together as a dualised theory.

\subsection{Combining the Dual and Dualised Standard Models}

In the previous section we discussed how the dual and dualised standard
models are complementary theories. This motivates that perhaps they
should be combined together to give a full description of the standard model.
A natural way to do this would be to dualise the dual standard model by
somehow inducing colour charges on the monopoles.

However as the two models presently stand there are difficulties with this
dualisation. This is because the construction of dual colour Yukawa couplings 
requires dual colour matter assignments ${\bm 3}_L, {\bm 1}_R\x 3$,
with only $(u,d)_L$ and $(\nu,e)_L$ dyonic~\cite{chandsm}. 
That is, the dualised standard model
derives three generations by postulating a dual colour structure analogous 
to electroweak isospin. The evidence for this are the rather accurate 
estimations of fermion masses and mixing angles~\cite{chanckm}.

However this structure is not compatible with dualising the dual standard 
model. There only one generation of $u_R$, $d_R$ and $e_R$ are derived; not 
the three required to construct dual colour Yukawa couplings. Instead,
from a dual standard model perspective, it appears natural that
all SU(5) monopoles should somehow gain dual colour charge; then
three generations would originate solely from dual colour. Certainly the
physical mechanism of dual colour breaking still appears to apply, although
the fermion masses and mixing angles would not.

Perhaps an investigation of the effective 
couplings between monopoles may yield similar couplings; for
instance if ${\bm 3}_R$ is first broken to ${\bm 1}_R\x 3$
then such Yukawa couplings can be constructed. In the dualised standard model 
these Yukawa couplings are effective anyway; since they are not
gauge invariant unless derived from a non-renormalisable 
interaction~\cite{chandsm}.

Perhaps many of these issues relate to quantising the dual
standard model (indeed we will see later there are other problems 
with quantisation). As a preliminary investigation one might determine
whether such dual colour charges may naturally occur within the classical 
monopole theory. That is the subject of the next section.

\section{Dualising the Dual Standard Model}
\label{dual}

In the previous section we motivated that perhaps one should 
combine the dual standard model and the dualised standard model together.
Within this section we give an illustration of the sort of methods that 
could be used.

Essentially we examine here only whether the most basic proposal in
the dualised standard model is consistent with a dual standard model. That
is whether the dual standard model can be naturally dualised such that
every elementary particle has the dual colour required for three generations.
In addition we check whether this procedure is consistent with the parity 
and angular momentum assignments of the elementary particles.

We should make it clear that there are some difficult issues 
when extrapolating the 
following to the fully quantised regime. These issues and some potential
resolutions are discussed in sec.~(\ref{problems}). For this reason 
the following represents an investigation of whether a consistent 
classical/semi-classical theory can be obtained. 

\subsection{SU(5) Monopoles in a Theta Vacuum}
\label{sec1}

To start we consider the effects of a theta vacuum on the SU(5)
monopole spectrum of sec.~(\ref{dsm}). Such a theta vacuum has been motivated
to play an important role in formulating a dual standard 
model~\cite{vach95,vachdsm}. This is because the SU(5) monopole spectrum is 
parity invariant (unlike the standard model) unless a theta vacuum is 
included.

The effect of a theta term in the SU(5) gauge theory
\beq
\label{ltheta}
\calL_\theta = \frac{\theta g^2}{8\pi^2}\,\tr\bm E \cdot \bm B
\eeq
is to induce theta dependent electric charge~\cite{witten} on the monopoles in 
table~\ref{tab1}. A simple way to see this is to consider 
the interaction of a monopole with a gauge field $(\phi,\bm a)$. 
Following an argument of Coleman's~\cite{cole} 
the electric and magnetic fields  
\beq
\bm E = \bm\nabla \phi,\hspace{1em}
\bm B = \bm\nabla \wedge \bm a + \frac{1}{2g} \frac{\rhat}{r^2}M, 
\eeq
are substituted into (\ref{ltheta}) to give, upon integration by parts,
\beq
L_\theta = \int {\rm d}^3\bm r\  \calL_\theta =
 - \frac{\theta g}{2\pi} \int {\rm d}^3\bm r\  \de^3(\bm r)\,\tr\, \phi M.
\eeq
But this is precisely the interaction between the
gauge potential and an electric charge $Q_\theta = - \ts\frac{\theta}{2\pi}M$.
Consequently each monopole in table~\ref{tab1} gains a theta
dependent electric charge, becoming a dyon
\beq
\label{thetae}
\bm E \sim -\frac{\theta g}{2\pi}\frac{\hat\bm r}{4\pi r^2}M,\hspace{1em}
\bm B \sim \frac{1}{2g}\frac{\hat\bm r}{r^2}M.
\eeq

Here we are particularly interested in the effects of this theta vacuum
on the interactions of these monopoles with electric charges.
To be specific we consider the bosons associated with the components
$H_i$, $i=1,..,5$, of a $\bm 5$ scalar field. These source a non-Abelian
electric field
\beq
\bm E = g \frac{\rhat}{4\pi r^2}Q_i,
\eeq
with $Q_i$ the electric generator, which has the form
\bea
\label{Qi}
Q_1&=&{\rm diag}(\fourfifth,-\fifth,-\fifth,-\fifth,-\fifth),\nn\\
\cdots && \hspace{3em}\cdots\hspace{3em} \\
Q_5&=&{\rm diag}(-\fifth,-\fifth,-\fifth,-\fifth,\fourfifth).\nn
\eea
Similarly the $\bar H$ bosons have electric generators $\bar{Q}_i$
of opposite sign.
Then $\{Q_1,Q_2,Q_3\}$ form a colour triplet and $\{Q_4,Q_5\}$ form 
an isospin doublet; each with the appropriate colour and isospin charges.

Note that this scalar field $H$ associated with the above bosons
is directly relevant 
to constructing a dual standard model. In points (ii) and (iii) of 
sec.~(\ref{dsm}) this field is used to obtain fermions from bosons 
and for breaking dual colour.

Because of the global properties of 
charge around a monopole the theta induced charge is always associated with
an Abelian interaction. That is, not all electric charges can be defined
in the presence of a monopole (with some gaining an infinite energy string
singularity), but the theta induced charge is necessarily well defined
as a $\U(1)_M$ interaction~\cite{globcol,bala}. This induced charge then 
interacts with the $H_i$ bosons through their $\U(1)_M$ charge components.

The feature we wish to draw attention to is that the theta induced charge will
attract some electric charges and repel others in a parity violating manner. 
Whether
they are attracted or repelled depends upon whether the $H_i$ bosons have 
positive/negative $\U(1)_M$ charges. This provides a natural,
parity violating mechanism for forming dyonic composites. Many of these have
one-half angular momentum, as is necessary to construct a dual standard model.

This can be illustrated by considering the theta induced Coulomb potential 
between a monopole with generator $M$ and a charge $Q$
\beq
\label{pot}
V(r) \sim -\frac{\theta g^2}{2\pi} \frac{\tr QM}{4\pi r},
\hspace{2em} r\gsim R_c.
\eeq
Here both the charge and monopole are approximated as point sources 
outside the monopole core. Inside the core the magnetic field (and hence
induced electric field) decreases continuously to zero by Gauss's law.
Clearly this potential is binding/repulsive depending upon whether
$\tr{QM}$ is positive/negative.
 
That the potential (\ref{pot}) results in a parity violating spectrum of 
bound dyons is because of the even/odd properties of the electric/magnetic 
fields under parity inversion. Then $\calP:(Q,M)\mapsto(Q,-M)$
takes a bound dyon into a non-bound state.

An interesting property of the stable dyons is that their angular momentum 
assignments
also violate parity. For scalar electric charges the angular momentum of the
resulting dyonic composites is
\beq
\label{J}
J_3 = \int {\rm d}^3r\, [\bm r \wedge (\bm E \wedge \bm B)]_3= \half\,\tr{QM},
\eeq
with the monopole-charge axis orientated to $\hat\bm x_3$. 
Then their parity conjugates have the opposite angular momentum.

\subsection{The Dual Standard Model in a Theta Vacuum}

We now apply the above properties of a theta vacuum to the 
construction of a dual standard model. The central idea is to use the theta
binding effect to naturally form a parity violating spectrum of SU(5)
dyons, all of which have one-half angular momenta.

As well as trying to achieve angular momentum assignments compatible with
the standard model we would also like the resulting dyon spectrum to be
compatible with obtaining three generations through the methods of Chan
and Tsou. To help along these lines we take some indications from
the dualised standard model. There they require dual colour to be broken,
whilst dual isospin appears to be confining, with a large confinement scale 
(say over a hundred GeV) to not be presently observed.

Therefore we do not consider SU(5) dyons with electric isospin, as 
in the dualised standard model these are confined into very heavy dual 
isospin hadrons. As we will see this conveniently simplifies
the following calculations.

Such composite dyons can be formed by combining the monopoles in 
table~\ref{tab1} with the charges $\{Q_1,Q_2,Q_3\}$ in (\ref{Qi}). However 
these are not the only dyons present in the dual standard model;
there are also non-Abelian analogues of the Julia-Zee dyon~\cite{julia75}. 
In principle the theta induced charge can also 
bind their electric charge to the monopoles.

The description of these monopole gauge excitations
is quite involved and we refer to ref.~\cite{comp} for a fuller
discussion. Care has to be taken with the global properties of electric charge,
because not all charges are well defined around a magnetic monopole.
The dyon spectrum is obtained upon performing a semi-classical quantisation
of the global electric degrees of freedom around a monopole. This results in 
the following spectrum of possible electric charges, with colour,
isospin and hypercharges defined through
$Q=q_\rmC T_\rmC + q_\rmI T_\rmI + q_\rmY T_\rmY$:

\begin{table}[h]
\caption{Gauge excitations of the monopoles.}
\begin{eqnarray*}
\begin{array}{|c|ccc|c|}
\hline
{\rm diag}\ Q &q_{\rm C}&q_{\rm I}&q_{\rm Y}&{\rm allowed\ on} \\
\hline
\vspace*{-0.35cm} &&&& \\
(0,0,1,$-$1,0)& 1 & \half & \third & {\rm all} \\
\vspace*{-0.35cm} &&&& \\
(0,1,1,$-$1,$-$1)& $-1$ & 0 & \twothird & \bar d, (\bar\nu, \bar e), u, 
\bar e\\ 
\vspace*{-0.35cm} &&&& \\
(1,1,1,$-$2,$-$1) & 0 & $-$\half & 1 & (\bar\nu, \bar e), u, \bar e\\
\vspace*{-0.35cm} &&&& \\
(1,1,2,$-$2,$-$2)& 1 & 0 & \ts\frac{4}{3} & (\bar\nu, \bar e), u, \bar e\\
\vspace*{-0.35cm} &&&& \\
(1,2,2,$-$3,$-$2)& $-$1 & \half & \ts\frac{5}{3} & (\bar\nu,\bar e), u, 
\bar e\\
\vspace*{-0.35cm} &&&& \\
(2,2,2,$-$3,$-$3)& 0 & 0 & 2 & (\bar\nu, \bar e), u, \bar e \\
\hline 
\end{array}
\end{eqnarray*}
\vspace*{-0.5cm}
\label{tab3}
\end{table}

\noindent
Here we restrict our attention to the lower charged and therefore least 
energetic excitations. Of these there are three that are 
uncharged under isospin
\bea
Q_\rmC &=& {\rm diag}(0,1,1,\minus 1,\minus 1),\hspace{2em}
Q_\rmCp = {\rm diag}(1,1,2,\minus 2,\minus 2),\nn\\
&&\hspace{3em}Q_\rmY = {\rm diag}(2,2,2,\minus 3,\minus 3).
\eea
From now on we will take these charges as input, and not consider
their origin from the semi-classical quantisation.

Now we need to determine the angular momenta of these gauge excited dyons. 
For this there are two situations:

\noindent (i) Spherically symmetric dyons have vanishing angular momentum. 
There is a simple criterion for determining whether the dyons are spherically
symmetric from their $(Q,M)$ charge, as described in sec.~(VIII) of 
ref.~\cite{comp}. 

\noindent (ii) Otherwise dyons will have angular momentum, although there
are many issues that have not been fully understood. As a simple model
for calculating their angular momentum we consider $Q$ to be composed of 
two components $Q=Q_0+Q_s$, where $Q_0$ defines a spherically symmetric dyon. 
Then the angular momentum originates from $Q_s$, which we interpret 
as a single gauge boson in the background of the monopole. This gives
\beq
\label{gex}
J_3 = \left\{ \begin{array}{c} \half\,\tr Q_sM - 1,\ \ \ \tr Q_sM\geq 0,\\
\half\,\tr Q_sM + 1,\ \ \ \tr Q_sM\leq 0, \end{array} \right.
\eeq
in which we have included the spin of the gauge boson as being energetically
orientated opposite to the magnetic field~\cite{comp} and taken the 
monopole-charge axis as $\bm x_3$.

It should be noted that there are many issues with gauge excitations 
that have not been fully understood. However the above configurations
are present in the SU(5) monopole theory, and one may use Goldhaber's
argument~\cite{gold:spin} to show they are fermionic. It is possible that 
(\ref{gex}) is only valid for some gauge
excitations, although we have checked that those dyons in table~\ref{tab4}
below are compatible with the semi-classical analysis of Dixon~\cite{dixon84}.

From this we can determine the angular momenta 
of the appropriate dyons in the dual standard model:

\begin{table}[h]
\caption{Angular momenta of the dyons.}
\begin{eqnarray*}
\begin{array}{|c|c|cccccc|}
\hline
n & {\rm multiplet} & Q_1 & Q_2 & Q_3 & Q_\rmC & Q_\rmC' & Q_\rmY \\
\hline
\vspace*{-0.35cm} &&&&&&& \\
1&(u,d) & 0 & 0 & \half & - & - & - \\
\vspace*{-0.35cm} &&&&&&& \\
2&\bar d & 0 & \half & \half & 0 & - & - \\ 
\vspace*{-0.35cm} &&&&&&& \\
3&(\bar\nu, \bar e) & \half & \half & \half  & -\half & \pm 1 &  0 \\
\vspace*{-0.35cm} &&&&&&& \\
4&u & \half & \half & 1 & -\half & 0 & 0 \\
\vspace*{-0.35cm} &&&&&&& \\
6&\bar e & 1 & 1 & 1 & \pm 1 & \pm 1 & 0 \\
\vspace*{-0.35cm} &&&&&&& \\
6&\bar e^* & \half & \half & 2 & \half & 0 & 0 \\
\hline 
\end{array}
\end{eqnarray*}
\vspace*{-0.5cm}
\label{tab4}
\end{table}
\noindent
In this table we also 
include an $\bar e^*$ monopole with $M=(1,1,4,\minus 3,\minus 3)$,
which has extra non-topological magnetic charge.
Vachaspati and Steer have motivated that this
non-topological degree of freedom relates to the internal structure
of the monopole, so that the long range magnetic interactions are
unaffected. For more details we refer to their paper~\cite{steer}.

For the scalar boson-monopole composites 
only the states $(Q,M)$ with non-zero angular momentum can be bound 
configurations. This is because both the binding potential (\ref{pot}) and
angular momentum (\ref{J}) are proportional to tr$\, QM$. The charge 
conjugates $(-Q,-M)$ have also the same stability and 
angular momenta, whilst the parity conjugates $(Q,-M)$ have
the opposite stabilities.

In conclusion this mechanism has been of some success within
the dual standard model. Certainly all of the monopoles become
naturally dyonic, with a spectrum of non-zero angular momenta that violates
parity maximally. Also all are colour charged, 
as is consistent with a dualised standard model. However there are problems:

\noindent (i)
There is no $\bar{e}$ dyon with one-half angular momentum. To
construct such a state requires two quanta of $H$, which is problematic as 
such a dyon possesses dual isospin.

\noindent (ii)
As well as the desired standard model states there are other angular momentum
analogues for the $(\bar\nu,\bar e)$, $u$ and $\bar{e}$ dyons. Certainly
no such states have been observed in the standard model.

A way of solving problem (i) is to instead consider the $\bar e^*$ monopole,
which has dyons with one-half angular momentum. Later we will see that
the $\bar e$ dyons have a complicated energy spectrum at
non-zero theta. This raises the possibility that a dyon with
non-topological magnetic charge could be the admissible  state.

Problem (ii) is less straightforward. It seems that some energetic 
criterion should be applied. We examine this in the next section.

\subsection{Dualising the Dual Standard Model}
\label{sec3}

As the above mechanism stands there is another reason why all
the above composites cannot represent
standard model fermions. One should require each composite's mass to be
less than their possible decay products, since only then are the dyons
absolutely stable to decay. Note that similar ideas have been 
proposed in ref.~\cite{steer}, although the following discussion
is very different from that.

In this section we will take the charges and spins of the dyons as
input and consider only the classical charge-monopole interactions.
This is because there are some difficult issues associated with a fully
quantised treatment, as discussed in the next section.

The SU(5) monopole masses can be estimated in a fairly simple way from their 
solitonic properties. Their scalar core energy and magnetic mass 
are determined by the coupling $g$, vev $v$, and core-size $R_c$
\bea
\calE_s &\sim& \half\ds\int_{r<R_c}{\rm d}^3r\,\tr\!\mod{\bm D \Phi}^2 
\sim 2\pi R_cv^2,\\
\calE_B &\sim& \half\ds\int_{r>R_c}{\rm d}^3r \,\tr B^2 
\sim \frac{2\pi}{g^2R_c}\mod{M}^2,
\eea
where $\mod{M}^2=\tr{M^2}$. For simplicity we approximate the scalar and 
gauge core sizes as equal, which does not appreciably effect the central 
result~(\ref{elen}) below.
An equilibrium between the scalar and magnetic energies leads to
\beq
\label{rm}
R_c \sim \frac{\mod{M}}{gv},\hspace{2em} 
m_{\rm mon}\sim \calE_B+\calE_s \sim \frac{4\pi v}{g}\mod{M}.
\eeq
It is interesting that the electromagnetic mass essentially
determines the monopole's mass $m_{\rm mon}$. 
The problem of electromagnetic mass has
a long history (see ref.~\cite{feyn}); for instance it diverges in many 
situations. For the above solitons the role of electromagnetic 
mass is clear: it simply constitutes half of the monopole's mass.

The electromagnetic mass is also central to calculating the
soliton's mass in a theta vacuum. Then the total mass is the sum of 
$m_{\rm mon}$ in (\ref{rm}) and the mass in the theta induced electric field
(\ref{thetae})
\beq
\calE_E \sim \half\ds\int_{r>R_c}{\rm d}^3r \,\tr E^2
\sim \frac{\theta^2g^3v}{32\pi^3}\mod{M}.
\eeq

An interesting point is that a dyon's electric charge 
can cancel off part of the theta induced electric field. 
This will decrease the electric mass of the dyon.
There are many issues with this observation, but let us 
explore the consequences for the stable dyon spectrum.

Considering a dyon $(Q,M)$ in a theta vacuum,
\beq
\bm E \sim g\frac{\rhat}{4\pi r^2}(Q-\frac{\theta}{2\pi}M),\hspace{2em}
\bm B \sim \frac{1}{2g}\frac{\rhat}{r^2}M.
\eeq
Then the electric mass of this dyon is 
\beq
\label{elen}
\calE_E \sim \half\ds\int_{r>R_c}{\rm d}^3r\,\tr E^2 
\sim \frac{g^3v}{8\pi\mod{M}}\,\tr(Q-\frac{\theta}{2\pi}M)^2,
\eeq
whilst the magnetic mass stays the same.

\begin{figure}[p]
\begin{center}
\epsfxsize=20em \epsfysize=15em \epsffile{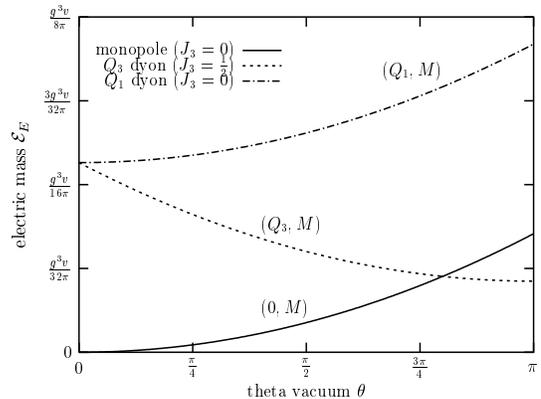}
\end{center}
\caption{Electric mass of the $(u,d)$ monopole and dyons $(Q_3,M)$ and 
$(Q_1,M)$. For $\theta\in(\frac{4}{5}\pi,\pi)$ the $(Q_3,M)$ dyon with
$J_3=\half$ has least electric mass.}
\label{fig2}
\end{figure}

\begin{figure}[p]
\begin{center}
\epsfxsize=20em \epsfysize=15em \epsffile{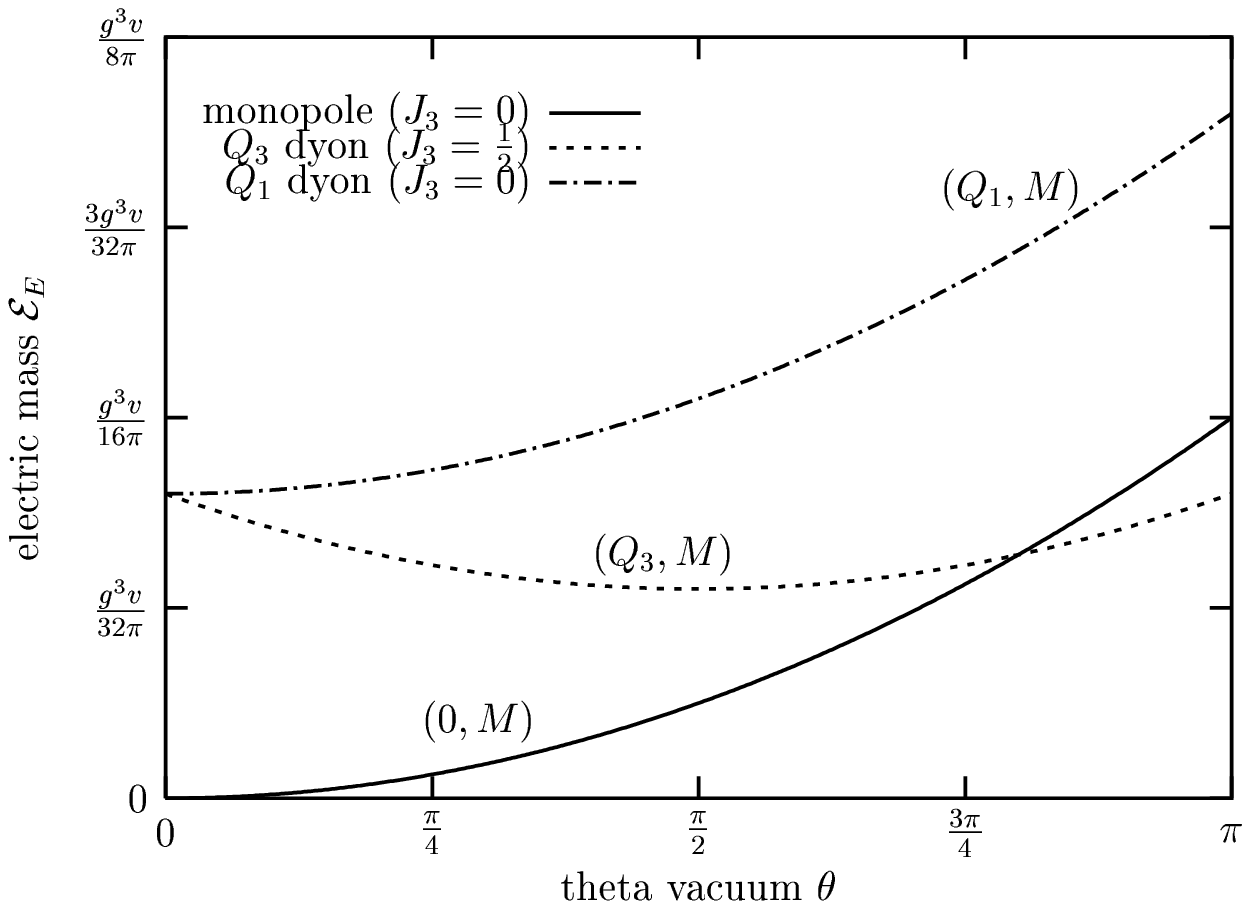}
\end{center}
\caption{Electric mass of the $\bar d$ monopole and dyons 
$(Q_3,M)$ and $(Q_1,M)$. For $\theta\in(\frac{4}{5}\pi,\pi)$ the
$(Q_3,M)$ dyon with $J_3=\half$ has least electric mass.}
\label{fig3}
\end{figure}

\begin{figure}[p]
\begin{center}
\epsfxsize=20em \epsfysize=15em \epsffile{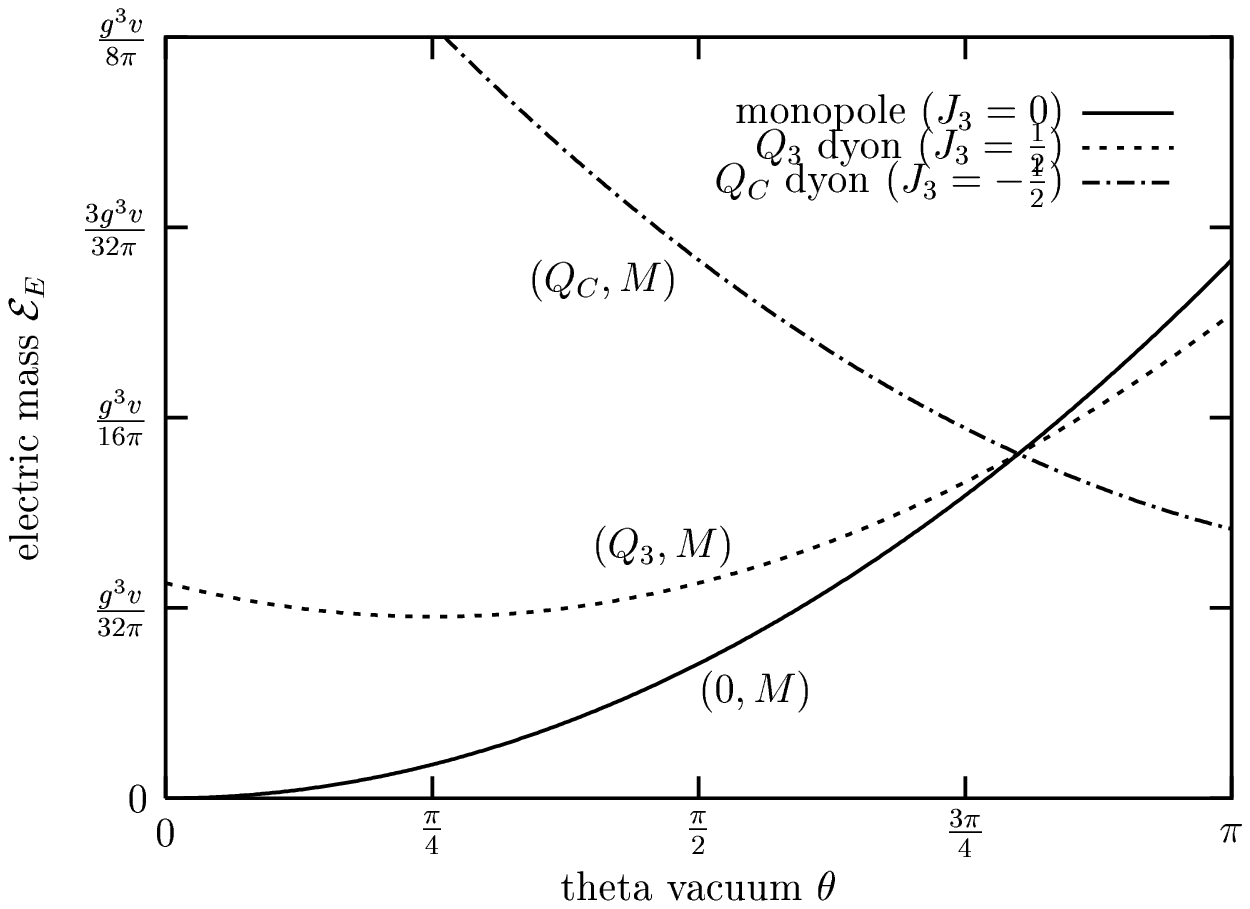}
\end{center}
\caption{Electric mass of the $(\bar\nu,\bar e)$ monopole and dyons 
$(Q_3,M)$ and $(Q_\rmC,M)$. For $\theta\in(\frac{4}{5}\pi,\pi)$ the 
$(Q_\rmC,M)$ dyon with $J_3=-\half$ has least electric mass.}
\label{fig4}
\end{figure}

\begin{figure}[p]
\begin{center}
\epsfxsize=20em \epsfysize=15em \epsffile{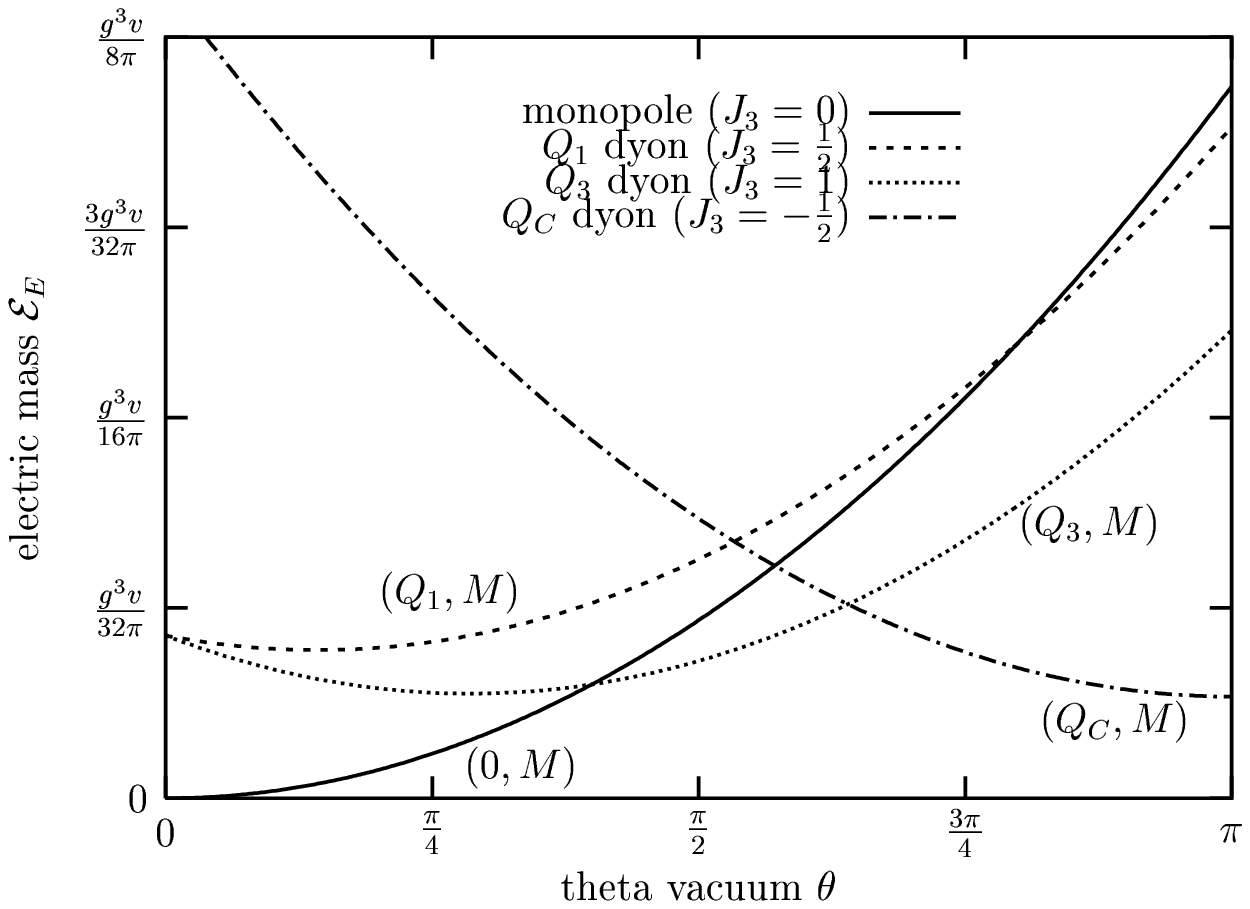}
\end{center}
\caption{Electric mass of the $u$ monopole and dyons $(Q_1,M)$, $(Q_3,M)$
and $(Q_\rmC,M)$. For a range of theta the
$(Q_\rmC,M)$ dyon with $J_3=-\half$ has least electric mass.}
\label{fig5}
\end{figure}

\begin{figure}[p]
\begin{center}
\epsfxsize=20em \epsfysize=15em \epsffile{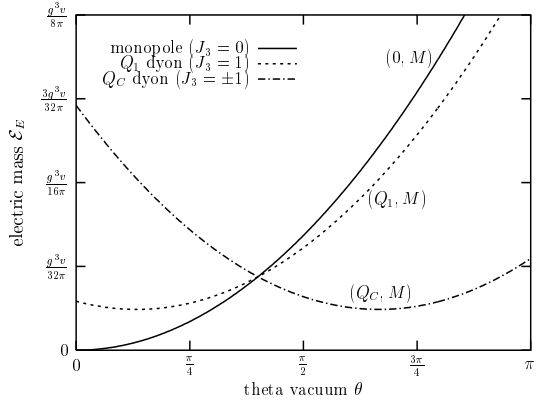}
\end{center}
\caption{Electric mass of the $\bar e$ monopole and the dyons $(Q_1,M)$ and 
$(Q_\rmC,M)$. For a range of theta the 
$(Q_\rmC,M)$ dyon with $J_3=1$ has least electric mass.}
\label{fig6}
\end{figure}

\begin{figure}[p]
\begin{center}
\epsfxsize=20em \epsfysize=15em \epsffile{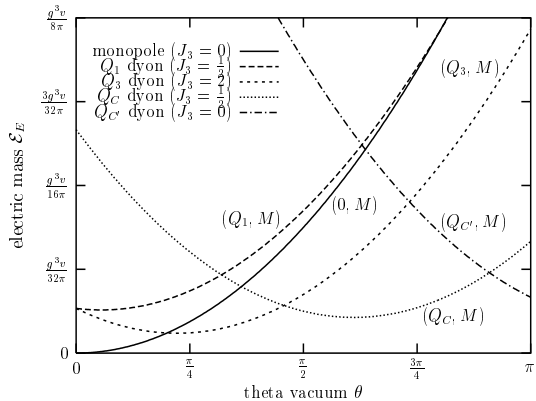}
\end{center}
\caption{Electric mass of the $\bar e^*$ monopole and the dyons $(Q_1,M)$,
$(Q_3,M)$, $(Q_{\rmC'},M)$ and $(Q_\rmC,M)$. For
$\theta\in(\frac{4}{5}\pi,\frac{10}{11}\pi)$ the $(Q_\rmC,M)$ 
dyon with $J_3=\half$ has least electric mass.}
\label{fig7}
\end{figure}

For the dyons in table~\ref{tab4} we now plot all of their electric energies
with theta in figs.~\ref{fig2} to \ref{fig7}. Those dyons not included on 
the figures, which includes the different gauge orientations, have been 
verified to not be of least energy. 

In conclusion for $\theta\in(\frac{4}{5}\pi,\frac{10}{11}\pi)$ the states with
least electric mass are: 

\begin{table}[h]
\caption{States of least electric mass for 
$\theta\in(\frac{4}{5}\pi,\frac{10}{11}\pi)$.}
\begin{eqnarray*}
\begin{array}{|c|c|c|c|c|}
\hline
n& {\rm diag}\ Q & {\rm diag}\ M & \ J_3\  & {\rm fermion} \\
\hline
\vspace*{-0.35cm} &&&& \\
1&($-$\fifth,$-$\fifth,\fourfifth,$-$\fifth,$-$\fifth)
& (0,0,1,$-$1,0) & \half &(u,d)_L \\ 
\vspace*{-0.35cm} &&&& \\
2&($-$\fifth,$-$\fifth,\fourfifth,$-$\fifth,$-$\fifth) & (0,1,1,$-$1,$-$1)&
\half & \bar{d}_L \\ 
\vspace*{-0.35cm} &&&& \\
3&(0,1,1,$-$1,$-$1) & (1,1,1,$-$1,$-$2) & $-$\half & (\bar\nu,\bar e)_R \\
\vspace*{-0.35cm} &&&& \\
4&(0,1,1,$-$1,$-$1) & (1,1,2,$-$2,$-$2) & $-$\half & u_R \\
\vspace*{-0.35cm} &&&& \\
6&(0,1,1,$-$1,$-$1) & (2,2,2,$-$3,$-$3) & 1 & - \\
\vspace*{-0.35cm} &&&& \\
6&(0,1,1,$-$1,$-$1) & (1,1,4,$-$2,$-$2) & \half & \bar{e}^*_L \\
\hline 
\end{array}
\end{eqnarray*}
\vspace*{-0.5cm}
\label{tab5}
\end{table}

\noindent
Thus if the criterion for selecting relevant states is
the dyon's electric mass then this does give the required spectrum, with
all states having $\mod{J_3}=\half$. We stress that in no way was this result
necessary or predetermined; the dynamics just happened to give the desired 
answer. There are difficulties with the $\bar e$ states, but $\bar e^*$ may
be less massive and these two monopoles differ only by their internal
structure~\cite{steer}.

Note that all dyons in table~\ref{tab5} are electrically colour charged, and
transform fundamentally under electric colour.
Therefore their dyonic charges are compatible with Chan and Tsou's 
interpretation of three generations.


\newpage
An interesting and unexpected bonus of the above calculation is that we also
appear to have obtained an association between the fermion's chirality
assignments and the sign of the dyon's angular momenta. In the high momentum
limit $p\gg m$
the angular momentum of a chiral fermion is unambiguously determined
through the helicity projection operator $\Pi^{\pm}(\bm p)=\half(1\pm\ga_5)$. 
In that limit its angular momentum along the direction of motion is
\beq
J_3\psi_L=\half\psi_L,\hspace{2em}J_3\psi_R=-\half\psi_R.
\eeq

Therefore table~\ref{tab5} also gives a correspondence between
these angular momenta and those of the associated dyons. Again
the pattern $\{+\half,+\half,-\half,-\half,+\half\}$ occurred 
through the specific dynamics of the situation.

Thus the question is: does this least electric mass criterion
justify that the dyon is stable? Fortunately there is at least one 
situation where this appears to be so.

At strong electric coupling $g^2/4\pi\gg 1$ in a theta vacuum 
a monopole's electric mass is the dominant mass contribution.
Then the magnetic mass, the gauge boson masses, and the scalar boson masses
may be consistently taken to be much smaller than this electric mass. 
If so then figs.~\ref{fig2} to \ref{fig7} appear to represent the dyon's 
masses, of which table~\ref{tab5} contains those with least mass.

However there are then some difficult issues, mainly relating
to quantisation.  These are discussed in the next section.

\subsection{Issues and Interpretation}
\label{problems}

In the previous section we saw that a reasonable spectrum of dyons could
be derived providing the gauge theory is strongly coupled. However
there are then some difficult issues, as we now explain.

When the gauge theory is strongly coupled the nature of the monopoles is
different from at weak coupling. This can be seen by comparing 
their core size $R_c$ with their Compton wavelength $\la\sim m^{-1}$
\beq
\label{22}
\la/R_c \sim g^2/4\pi.
\eeq
Thus at weak coupling the monopoles are classical soliton configurations,
whilst at strong coupling they are fully quantum mechanical.

At first sight this seems promising for the dual standard model because the
observed fermions are fully quantum mechanical. That is, the observed 
elementary
particles are not quantised semi-classically but are fully quantised through
methods such as the path-integral formalism. Then it would seem as if a proper
quantisation of the dual standard model should yield a quantum field theory
similar to the standard model. 

However some difficult issues arise when extending the calculations in 
secs.~(\ref{sec1}-\ref{sec3}) to the fully quantum regime at strong
coupling. Specifically:

\noindent (i) Whilst the theta induced charge is valid at strong coupling
in a classical context, the effects of quantisation are not known.

\noindent (ii) Also, a substantial theta vacuum would give a strong 
$\calC\calP$ problem. The strong interactions are time reversal symmetric
to large accuracy and a large $\theta$ would appear to be at odds with
this. Additionally, although we have not discussed weak $\calC\calP$
violation in this work, the induced violation would be too large there as 
well.

It is beyond the scope of this paper to fully 
address these issues. Indeed, as we
discuss below, their resolution may require a detailed understanding of 
the quantisation of solitons at strong coupling; a subject that is poorly
understood. We discuss now a couple of different perspectives on these
issues.

Firstly it could transpire that the dual standard model is only appropriate
as a classical, effective description of the elementary particles. This
would be analogous to the Skyrme model, in which baryons are 
accurately described as solitons of a classical field theory~\cite{skyrme}. 

It is worth commenting that the modern interpretation of the Skyrme model 
is as a consequence of large $N$ QCD~\cite{largeN}, 
whereas the dual standard model
is motivated by SU(5) unification and duality. Otherwise the 
Skyrme model and dual standard model share many similar features and outlook;
indeed, many of Skyrme's original motivations also apply to the dual standard
model~\cite{sky88}.

A second perspective is to tackle the above issues through a full 
quantisation of the dual standard model. Unfortunately, the techniques
to carry out such a program have not been developed, so that
no definitive conclusions can presently be made. 
However some tentative suggestions for resolving (i) and (ii) might then be:

\noindent (i) Perhaps the stable dyons can be represented by a second
quantised fermionic field theory. In this sense the field theory may
be largely insensitive to the internal details of the dyon, somewhat in
analogy to how proton-electromagnetic interactions are described by QED, 
even though the proton has a quark substructure. 

\noindent (ii) If a full quantisation of the SU(5) solitons was to yield a
reasonable description of the standard model fermions then an $O(1)$ 
contribution to the theta angle should arise from the determinant of the
resulting quark mass matrix. Perhaps
this could be arranged to cancel with the classical part used in this paper. 
In some sense this situation arises naturally in the standard model, since
a suitable $\theta F\tilde{F}$ term is required to cancel this contribution; 
however here we interpret this necessary theta term 
as a starting ingredient of the dual standard model.

Whatever the interpretation we wish to stress that the classical
arguments given in this paper do appear to produce the desired angular
momentum assignments of the elementary particles. This is on top of the
other successful features of the dual standard model.

\section{Conclusion}
\label{conc}

In this paper we have stressed that the dual and dualised standard models
are complementary theories. That is, their implications have no overlap, 
whilst
taking their consequences together appears to yield an explanation for 
most of the standard model. For instance the dual standard model
explains the properties of one generation of standard model 
fermions as solitons originating from SU(5) gauge unification; whilst the 
dualised standard model explains the properties of three 
generations as originating from a dualised fermion spectrum.

A central aspect of this paper is the suggestion that inclusion of a 
theta vacuum naturally combines these two theories together. 
Such a theta vacuum is expected to be important for a dual
standard model because it should play a role in introducing parity violation. 
We have shown that in addition to this it has the effect of dualising the 
soliton spectrum. In doing so it also suggests an explanation
for the chirality assignments of the elementary particles; a feature that
cannot be derived in either of the original dual or dualised standard models.

There are difficult issues associated with the interpretation of these 
calculations at strong coupling. Whilst the arguments are well motivated 
classically, it is beyond the scope of this paper to carry them over to the 
fully quantised regime. We note, however, that the methodology is natural and 
the conclusions are consistent with the standard model.

If a consistent quantisation of the dual standard model does allows it to be
naturally dualised by a theta vacuum, then perhaps a 
very simple theory of unification may ensue. In that instance particle and
gauge unification could consist of merely a broken dual theta-gauge theory
$\wt\SU(5)\rightarrow\wt\SU(3)_\rmC\x\wt\SU(2)_\rmI\x\wt\U(1)_\rmY/\bbZ_6$,
with {\em all} presently observed particle properties occurring within the 
resulting soliton spectrum.

\acknowledgments

Part of this work was supported by a junior research fellowship at
King's College, Cambridge. I thank H-M.~Chan, T.~Kibble, D.~Steer and  
S.~Tsou for help and advice. I am also indebted to T. ~Vachaspati
for his many helpful comments regarding this work.

%%%%%%%%%%%%%%%%%%%%%%%%%%%%%%%%%%%%%%%%%%%%%%%%%%%%%%%

\end{document}